\documentclass[12pt]{article}
\usepackage{epic}
\def\option{\noindent}
\def\vs{\vspace}
\newcommand{\Z}{\mathsf{Z}\kern -5pt \mathsf{Z}}
\def\half{ {1\over 2} }
\def\smhalf{  {\textstyle{1\over 2} }}
\def\id{ 0 } 
\def\a{\alpha}
\def\b{\beta}
\def\Lam{\Lambda}
\def\lam{\lambda}
\def\sig{\sigma}
\def\om{\omega}
\def\Pplus{ P_+^K }
\def\Spr{{S'}}
\def\Exp{  {\cal E}^{\om}  }
\def\Bound{  {\cal B}^{\om}  }
\def\Boundeps{  {\cal B}^{\varepsilon}  }
\def\ta{{\tilde{\a}}}
\def\tn{{\tilde{n}}}
\def\tb{{\tilde{\b}}}
\def\tlam{{\tilde{\lambda}}}
\def\tmu{{\tilde{\mu}}}
\def\tA{{\tilde{A}}}
\def\tB{{\tilde{B}}}
\def\tM{{\widetilde{M}}}
\def\tLam{{\widetilde{\Lambda}}}
\def\tS{{\tilde{S}}}
\def\tSpr{{\tS'}}
\def\ha{{\hat{\a}}}
\def\hb{{\hat{\b}}}
\def\hmu{{\hat{\mu}}}
\def\bz{{\bar z}}
\def\bV{\overline{V}}
\def\bJ{\overline{J}}
\def\bJ{\overline{J}}
\def\sothree{{{\rm so}(3)}}

\def\sooddn{{{\rm so}(2n+1)}}
\def\soevenn{{{\rm so}(2n)}}
\def\gcup{ {\breve g} }
\def\hgK{{\hat{g}_K}}   
\def\twhgK{{\hat{g}_K^\om}}   
\def\suNK{{\widehat{\rm su}(N)_K}}
\def\suKN{{\widehat{\rm su}(K)_N}}
\def\spnk{{\widehat{\rm sp}(n)_k}}
\def\spkn{{\widehat{\rm sp}(k)_n}}
\def\soNK{{\widehat{\rm so}(N)_K}}
\def\soKN{{\widehat{\rm so}(K)_N}}
\def\uNKh{{   \widehat{\rm u}(N)_{K,N(K+N)} }}
\def\uKNh{{   \widehat{\rm u}(K)_{N,K(K+N)} }}
\def\sothreeK{{\widehat{\rm so}(3)_K}}
\def\sofourK{{\widehat{\rm so}(4)_K}}
\def\sutwotwoK{{\widehat{\rm su}(2)_{2K}}}
\def\sooddnK{{\widehat{\rm so}(2n+1)_K}}
\def\sooddnoddk{{\widehat{\rm so}(2n+1)_{2k+1}}}
\def\sooddkoddn{{\widehat{\rm so}(2k+1)_{2n+1}}}
\def\soevennK{{\widehat{\rm so}(2n)_K}}
\def\soevennevenk{{\widehat{\rm so}(2n)_{2k}}}
\def\soevenkevenn{{\widehat{\rm so}(2k)_{2n}}}
\def\soevennoddk{{\widehat{\rm so}(2n)_{2k+1}}}
\def\sooddkevenn{{\widehat{\rm so}(2k+1)_{2n}}}
\def\somp{{\widehat{\rm so}(2n-1)_{K+1}}}
\def\sonmkp{{\widehat{\rm so}(2n-1)_{2k+1}}}
\def\sokmnp{{\widehat{\rm so}(2k-1)_{2n+1}}}
\def\DnoneK{  (D_{n}^{(1)})_K}
\def\DntwoK{  (D_{n}^{(2)})_K}
\def\twbdy{  | B \rangle\!\rangle^\om }
\def\unishimu{  | \mu \rangle\!\rangle_I }
\def\ishimu{  | \mu \rangle\!\rangle_I^\om }
\def\epsishimu{  | \mu \rangle\!\rangle_I^\varepsilon }
\def\uncardya{  | \a \rangle\!\rangle_C }
\def\omcardya{  | \a \rangle\!\rangle_C^\om }
\def\epscardya{  | \a \rangle\!\rangle_C^\varepsilon }
\def\equiva{   | \;[\a]\; \rangle\!\rangle }
\def\equivb{   | \;[\b]\; \rangle\!\rangle }
\def\equivta{   | \;[\ta]\; \rangle\!\rangle }
\def\equivha{   | \;[\ha]\; \rangle\!\rangle }
\def\nblama{ {n_{\b\lam}}^{\a}  }
\def\nbLama{ {n_{\b[\lam]}}^{\a}  }
\def\sblama{    {n_{[\b] [\lam] }^{~~~~~[\a]}  }}
\def\be{\begin{equation}}
\def\ee{\end{equation}}
\def\bea{\begin{eqnarray}}
\def\eea{\end{eqnarray}}

\begin{document}
\bibliographystyle{bst}

\begin{flushright}
BOW-PH-140
\end{flushright}
\vspace{30mm}

\vspace*{.3in}

\begin{center}
{\Large\bf\sf  
Level-rank duality of untwisted and twisted D-branes 	  \\
of the $\soNK$ WZW model\footnote{Research supported in part
by the NSF under grant PHY-0456944
}
}
\vskip 5mm Stephen G. Naculich\footnote{
\tt  naculich@bowdoin.edu}
and Benjamin H. Ripman
\end{center}

\begin{center}
{\em Department of Physics\\
Bowdoin College\\
Brunswick, ME 04011}

\end{center}
\vskip 2mm

\begin{abstract}

We analyze the level-rank duality of 
untwisted and $\varepsilon$-twisted D-branes of the $\soNK$ 
WZW model.
Untwisted D-branes of $\soNK$ are characterized by 
integrable tensor and spinor representations of $\soNK$.
Level-rank duality maps untwisted $\soNK$ D-branes 
corresponding to (equivalence classes of) 
tensor representations onto those of $\soKN$.
The $\varepsilon$-twisted D-branes of $\soevennevenk$ 
are characterized by (a subset of) integrable 
tensor and spinor representations of $\sonmkp$.
Level-rank duality maps spinor $\varepsilon$-twisted
$\soevennevenk$ D-branes onto those of $\soevenkevenn$.
For both untwisted and $\varepsilon$-twisted D-branes,
we prove that the spectrum of an open string 
ending on these D-branes is isomorphic to 
the spectrum of an open string 
ending on the level-rank-dual D-branes.

\end{abstract}

\vfil\break

%\renewcommand{\baselinestretch}{2}
%\small\normalsize

\section{Introduction}
\setcounter{equation}{0}
\label{secintro}

It has long been known that the 
modular transformation matrix and fusion algebra 
of the Wess-Zumino-Witten (WZW) model 
with affine Lie algebra $\suNK$ 
are closely related to those of the WZW model 
with affine Lie algebra $\suKN$
(level-rank duality)
\cite{Naculich:1990hg,%Naculich:1990bu,Fuchs:1989rv,Altschuler:1989nm,Walton:1988bs,Saleur:1990wv, Kuniba:1990im,Nakanishi:1990hj,
Naculich:1990pa, 
Mlawer:1990uv}.
Similar dualities have been shown for
WZW models with affine Lie algebras related by
$\spnk \leftrightarrow \spkn$  
and $\soNK \leftrightarrow \soKN$  \cite{
Naculich:1990pa, 
Mlawer:1990uv},
and also for $\uNKh \leftrightarrow \uKNh$ \cite{Naculich:2007nc}.

More recently,  it has been shown \cite{
Naculich:2005tn,
Naculich:2006mt, 
Naculich:2006bf}
that the untwisted and twisted D-branes 
in the boundary $\suNK$ WZW model
\cite{Klimcik:1996hp}--\cite{Gaberdiel:2004yn}
% %,Kato:1996nu, Alekseev:1998mc, Gawedzki:1999bq,
%Behrend:1998fd, %Behrend:1999bn, Fuchs:1999zi,
%Birke:1999ik,
%Alekseev:1999bs, %Alekseev:2000fd, Alekseev:2000jx, Alekseev:2000wg,
%Felder:1999ka,
%Stanciu:1999id, %Stanciu:2000fz, Stanciu:2001vw, Figueroa-O'Farrill:2000kz,
%Bachas:2000ik, %Pawelczyk:2000ah, Taylor:2000za,
%Fredenhagen:2000ei,
%Maldacena:2001ky,% Maldacena:2001xj,
%Gawedzki:2001ye, %Gawedzki:2002se,
%Ishikawa:2001zu,
%Petkova:2002yj,
%Gaberdiel:2002qa,
%Gaberdiel:2003kv,% Gaberdiel:2004hs,  Gaberdiel:2004za,
%Alekseev:2002rj, %Quella:2002wh, Quella:2002ct, Quella:2002ns,
%Bouwknegt:2002bq,% Bouwknegt:2003uj,
%Ishikawa:2002wx, % Ishikawa:2003xh, Ishikawa:2005ea, Ishikawa:2003kk, %Schafer-Nameki:2003aj, Braun:2004qg,
% Fredenhagen:2004xp, Vasudevan:2005ci, Fredenhagen:2005cj, Fredenhagen:2005an
respect level-rank duality;
that is,
there exists a one-to-one map between the (equivalence classes of) 
D-branes of $\suNK$ and those of $\suKN$. 
The open-string spectra associated with level-rank-dual 
D-branes are isomorphic,
and the charges of level-rank-dual 
untwisted D-branes are equal (modulo sign),
with a slightly more complicated relationship holding between
the charges of twisted D-branes.
Level-rank duality also holds for the
D-branes of $\spnk$ \cite{Naculich:2006mt}.

In this paper, we continue the story by establishing 
the level-rank duality 
of the untwisted D-branes of $\soNK$ and 
of the twisted D-branes of $\soevennevenk$.
In this case, level-rank duality is partial and holds only
for a subset of the D-branes of the theory.
Moreover, in this case we find no simple relation between
the charges of level-rank-dual D-branes.

We begin by summarizing our results.
Untwisted D-branes of $\soNK$ correspond to untwisted Cardy states $\uncardya$
(boundary states of the bulk WZW model), which are labelled by 
integrable highest-weight representations $\a$ (both tensors and spinors) 
of the untwisted affine Lie algebra $\soNK$.
Only untwisted {\it tensor} D-branes exhibit level-rank 
duality,\footnote{Except for 
$\sooddnoddk$, where a level-rank map can also be defined between
equivalence classes of untwisted spinor D-branes.} 
and the duality is one-to-one between 
equivalence classes $[\a]$ of integrable tensor representations 
generated by the $\Z_2$-automorphisms $\sig$ 
and (when $N$ is even) $\varepsilon$
of the $\soNK$ algebra.
(We denote by $\sig$ the simple current symmetry of $\soNK$ 
that acts on the Dynkin indices 
of a representation by\footnote{Except 
for $\sofourK$, in which case $\sig$ acts by
$a_0 \leftrightarrow \min(a_1,a_2)$.}
$a_0 \leftrightarrow a_1$.
We denote by $\varepsilon$ 
the ``chirality-flip'' symmetry of $\soevennK$ 
that acts on the Dynkin indices 
of a representation by $a_n \leftrightarrow a_{n-1}$.
For $\sooddnK$, we define $\varepsilon$ to be the identity.)
The boundary state corresponding to the equivalence class $[\a]$  
may be written as 
\be
\label{eq:ec}
\equiva
= 
{1 \over {\sqrt2}^{~t(\a) - s(\a) + 3 }}
\biggl[
  | \a \rangle\!\rangle_C
+ | \sig(\a) \rangle\!\rangle_C
+ | \varepsilon(\a)  \rangle\!\rangle_C
+ | \varepsilon(\sig(\a))  \rangle\!\rangle_C
\biggr]
\ee
where
\be 
s(\a) =  \left\{   
\begin{array}{ll}
1 &
{\rm if~} \a \neq \varepsilon(\a),  \\ [.05in]
0 &
{\rm if~}  \a =  \varepsilon(\a),
\end{array}
\right.
\qquad
t(\a) =  \left\{   
\begin{array}{ll}
1 &
{\rm if~} \a = \sig(\a),  \\ [.05in]
0 & 
{\rm if~} \a \neq \sig(\a). 
\end{array}
\right.
\ee
Equivalence classes $[\a]$ of integrable tensor representations
of $\soNK$ are characterized by Young tableaux with 
$ \le\!\! N/2$ rows and $ \le\!\! K/2$ columns.
Level-rank duality acts by transposing these tableaux,
inducing a one-to-one correspondence $[\a] \to [\ta]$
between equivalence classes of $\soNK$ and $\soKN$,
and therefore between the untwisted D-branes 
that correspond to the boundary states (\ref{eq:ec}).
We show that the spectrum of representations
carried by an open string stretched between untwisted
$\soNK$ D-branes corresponding to $[\a]$ and $[\b]$
is isomorphic to that
carried by an open string stretched between untwisted
$\soKN$ D-branes corresponding to $[\ta]$ and $[\tb]$.

The $\soevennK$ WZW model contains, in addition to untwisted D-branes,
a class of D-branes twisted by the symmetry $\varepsilon$.
These $\varepsilon$-twisted D-branes can be 
characterized \cite{Gaberdiel:2002qa}
by (a subset of) the integrable highest-weight representations 
(both tensors and spinors) of the 
untwisted affine Lie algebra $\somp$.
Only {\it spinor } $\varepsilon$-twisted D-branes 
of $\soevennevenk$ exhibit level-rank duality,
which involves a one-to-one map $\a \to \ha$ between 
the spinor $\varepsilon$-twisted D-branes of $\soevennevenk$ and 
the spinor $\varepsilon$-twisted D-branes of $\soevenkevenn$.
We show that the spectrum of representations
carried by an open string stretched between $\varepsilon$-twisted
$\soevennevenk$ D-branes corresponding to $\a$ and $\b$
is isomorphic to that
carried by an open string stretched between $\varepsilon$-twisted
$\soevenkevenn$ D-branes corresponding to $\ha$ and $\hb $.

This paper is organized as follows.
Section \ref{secsecond} briefly reviews
the Ishibashi and Cardy states of the WZW model,
and in sec.~\ref{secsoNK}, we characterize the
integrable highest-weight representations of $\soNK$.
Section \ref{secuntwisted} describes the level-rank
duality of the (equivalence classes of) 
untwisted D-branes corresponding to tensor representations of $\soNK$
and to spinor representations of $\sooddnoddk$.
The $\varepsilon$-twisted
Ishibashi and Cardy states of $\soevennK$
are reviewed in sec.~\ref{sectwisted}, 
and in sec.~\ref{sectwistedduality}
we describe the level-rank duality of 
spinor $\varepsilon$-twisted D-branes of $\soevennevenk$.

\section{Untwisted and twisted D-branes of WZW models}
\setcounter{equation}{0}
\label{secsecond}

In this section, we briefly review some general aspects of untwisted and 
twisted D-branes of the WZW model
and their relation to the Cardy and Ishibashi states 
of the closed-string sector, drawing on 
refs.~\cite{
Behrend:1998fd,
Birke:1999ik,
Ishikawa:2001zu,
Gaberdiel:2002qa}.

The WZW model,
which describes strings propagating on a group manifold,
is a rational conformal field theory
whose chiral algebra (for both left- and right-movers)
is an untwisted affine Lie algebra $\hgK$ at level $K$. 
We only consider WZW theories with a diagonal closed-string spectrum:
\be
\label{eq:diagonal}
{\cal H}^{\rm closed}  
= \bigoplus_{\lam \in \Pplus}  
 V_\lam \otimes \bV_{\lam^*}
\ee
where $V$ and $\bV$ represent left- and right-moving states
respectively,
$\lam^*$ denotes the representation conjugate to $\lam$,
and $\Pplus$ is the set of integrable highest-weight 
representations of $\hgK$. 

D-branes of the WZW model may be described algebraically 
in terms of the possible boundary conditions 
that can consistently be imposed on a WZW model with boundary.
We only consider boundary conditions 
on the currents of the affine Lie algebra of the form 
\be
\label{eq:twconditions}
\left[ J^a(z) - \omega \bJ^a(\bz)\right] \bigg|_{z=\bz} = 0
\ee
where $\omega$ is an automorphism of the Lie algebra $g$.
These boundary conditions leave unbroken the $\hgK$ symmetry, 
as well as the conformal symmetry, of the theory.

\vfil\break
\vs{.1in}
\noindent{\bf Twisted Ishibashi states.} 
\vs{.1in}

\option
Open-closed string duality allows one to correlate 
the boundary conditions (\ref{eq:twconditions}) 
of the boundary WZW model 
with coherent states $\twbdy \in {\cal H}^{\rm closed}$ 
of the bulk WZW model satisfying
\be
\label{eq:twmodes}
\left[  J^a_m + \omega \bJ^a_{-m} \right] \twbdy = 0\,, \qquad m\in \Z
\ee
where $J^a_m$ are the modes of the affine Lie algebra generators.
Solutions of eq.~(\ref{eq:twmodes})
that belong to a single sector 
$ V_\mu \otimes \bV_{\om(\mu)^*} $
of the bulk WZW model
are known as $\om$-twisted Ishibashi states 
$\ishimu$ \cite{Ishibashi:1988kg}.
As we are considering the diagonal closed-string theory
(\ref{eq:diagonal}),
$\om$-twisted Ishibashi states
only exist when $\mu = \om(\mu)$,
and so are labelled by
$\mu \in \Exp$, where $\Exp$ is the subset of
integrable representations of $\hgK$ 
that satisfy $\om(\mu)=\mu$.
Equivalently, $\mu$ corresponds to an integrable representation
of $\gcup^\om$,  
the orbit Lie algebra associated with $\hgK$ \cite{Fuchs:1995zr}.

\vs{.1in}
\noindent{\bf Twisted Cardy states.} 
\vs{.1in}

\option
A coherent state $\twbdy$ that corresponds to an
allowed boundary condition
must also satisfy additional (Cardy) conditions \cite{Cardy:1989ir}.
Solutions of eq.~(\ref{eq:twmodes}) that satisfy the
Cardy conditions are denoted $\om$-twisted Cardy states $\omcardya$,
where the labels $\a$ take values in $\Bound$,
the set of integrable representations 
of the $\om$-twisted affine Lie algebra 
$\twhgK$ \cite{Birke:1999ik}.
The $\om$-twisted Cardy states may be expressed 
as linear combinations of $\om$-twisted Ishibashi states
\be
\label{eq:defpsi}
\omcardya = \sum_{\mu \in \Exp}
{\psi_{\a\mu} \over \sqrt{S_{\id\mu}}} \ishimu 
\ee
where $S_{\lam\mu}$ is the modular transformation matrix of $\hgK$,
$\id$ denotes the identity representation,
and the coefficients $\psi_{\a\mu}$ may be identified 
with the modular transformation matrices 
of the $\om$-twisted affine Lie algebra $\twhgK$ \cite{Birke:1999ik}.

The $\om$-twisted D-branes of $\hgK$
correspond to the $\om$-twisted Cardy states $\omcardya$
and are therefore also labelled by $\a \in \Bound$.
The spectrum of an open string stretched between 
$\om$-twisted D-branes labelled by $\a$ and $\b$
is encoded in the open-string partition function 
\be
\label{eq:openpartition}
Z^{\rm open}_{\a\b} (\tau) 
= \sum_{\lam \in \Pplus} \nblama \chi_\lam (\tau) 
\ee
where $\chi_\lam (\tau)$
is the affine character of the integrable highest-weight 
representation $\lam$ of $\hgK$.
The multiplicity $\nblama$ of the representation $\lam$ 
carried by the open string 
may be expressed as \cite{Gaberdiel:2002qa}
\be
\label{eq:defn}
\nblama =  \sum_{\mu \in \Exp}
{ \psi^*_{\a\mu} S_{\lam\mu} \psi_{\b\mu} \over S_{\id\mu} }  \,.
\ee

\vs{.1in}
\noindent{\bf Untwisted Ishibashi and Cardy states.} 
\vs{.1in}

\option
Untwisted Cardy states 
$\uncardya$
and untwisted Ishibashi states 
$\unishimu$ 
are solutions of eq.~(\ref{eq:twmodes}) with $\om=1$,
and both are labelled by
integrable representations of $\hgK$.
The matrix 
$\psi_{\a\mu}$ 
in eq.~(\ref{eq:defpsi})
relating the untwisted Cardy states to the untwisted Ishibashi states 
is given by 
the modular transformation matrix 
$S_{\a\mu}$ of $\hgK$ \cite{Cardy:1989ir}.
Consequently,
by virtue of eq.~(\ref{eq:defn})
and the Verlinde formula for the fusion coefficients \cite{Verlinde:1988sn}
\be
\label{eq:verlinde}
\nblama 
=  
\sum_{\mu \in \Pplus}
{ S_{\b\mu} S_{\lam\mu} S^*_{\a\mu} \over S_{\id\mu} } 
=
{N_{\b\lam}}^{\a} 
\ee
the multiplicities $\nblama$
of the representations carried by an open string stretched 
between two untwisted D-branes $\a$ and $\b$
are given by the fusion coefficients ${N_{\b\lam}}^{\a} $
of the WZW model.

\section{Integrable representations of $\soNK$ }
\setcounter{equation}{0}
\label{secsoNK}

In this section, we review some
details about the integrable representations of $\soNK$ 
used throughout this paper.\footnote{Throughout this paper,
$N \ge 3$ is understood.}
Integrable representations of an affine Lie algebra $\hgK$ 
have non-negative Dynkin indices 
$(a_0, a_1, \cdots, a_{r})$
that satisfy $\sum_{i=0}^r m_i a_i = K$,
where $m_i$ are the dual Coxeter labels of the 
Dynkin diagram for $\hgK$,
and $r+1$ is the rank of $\hgK$.

\vs{.1in}
\noindent{\bf Integrable representations of $\sooddnK$.} 
\vs{.1in}

\option
The Dynkin diagram for $\sooddnK$ is

\begin{picture}(500,80)(10,10)
\put(100,40){\circle{5}}
\put(98,25){1}
\put(102,40){\line(1,0){26}}

\put(130,70){\circle{5}}
\put(120,66){0}
\put(130,68){\line(0,-1){26}}

\put(130,40){\circle{5}}
\put(128,25){2}
\put(132,40){\line(1,0){26}}

\put(160,40){\circle{5}}
\put(158,25){3}
\put(162,40){\line(1,0){26}}

\put(190,40){\circle{5}}
\put(188,25){4}
\put(192,40){\line(1,0){26}}

\put(220,40){\circle{5}}
\put(218,25){5}
\put(222,40){\line(1,0){26}}

\put(250,40){\circle{5}}
\put(248,25){6}
\put(252,40){\line(1,0){4}}
\put(258,40){\line(1,0){3}}
\put(263,40){\line(1,0){3}}
\put(268,40){\line(1,0){3}}
\put(273,40){\line(1,0){5}}

\put(280,40){\circle{5}}
\put(270,25){$n\!-\!2$}
\put(282,40){\line(1,0){26}}

\put(310,40){\circle{5}}
\put(301,25){$n\!-\!1$}
\put(312,41){\line(1,0){26}}
\drawline(327,40)(322,45)
\drawline(327,40)(322,35)
\put(312,39){\line(1,0){26}}

\put(340,40){\circle{5}}
\put(338,25){$n$}

\end{picture}
\option
and the dual Coxeter labels are 
($m_0$, $m_1$, $m_2, \cdots, m_{n-1}, m_{n}$)
= (1, 1, 2, $\cdots$, 2, 1),
where the labelling of nodes is indicated on the diagram.
Integrable representations of $\sooddnK$ thus have Dynkin indices that 
satisfy\footnote{
\label{fn}
Throughout this paper, by $\sothreeK$ we mean the affine
Lie algebra $\sutwotwoK$.
Its integrable representations have $\sothree$ Young tableaux 
that obey $\ell_1 \le K$.
Since $\ell_1 = \half a_1$, this means that eq.~(\ref{eq:sooddintegrable})
is replaced by  $a_0 + a_1 = 2K$ when $n=1$.}
\be
\label{eq:sooddintegrable}
a_0 + a_1 + 2(a_2 + \cdots + a_{n-1}) + a_n = K\,.
\ee
An even or odd value of $a_n$ corresponds, respectively,
to a tensor or spinor representation  of $\sooddn$.
With each irreducible tensor representation of $\sooddn$ 
may be associated a Young tableau 
whose row lengths $\ell_i$ are given by
\be
\label{eq:sooddrowlengths}
\ell_i = \left\{  
\begin{array}{ll}
\smhalf a_n + \sum_{j=i}^{n-1} a_j  
&
{\rm for~} 1 \le i \le n-1,  \\ [.05in]
\smhalf a_n 
&
{\rm for~} i=n.
\end{array}
\right. 
\ee
The integrability condition (\ref{eq:sooddintegrable}) 
is equivalent to the constraint 
$\ell_1 + \ell_2 \le K$ 
on the row lengths of the tableau.

We may also formally use eq.~(\ref{eq:sooddrowlengths}) 
to define row lengths for a spinor representation.
These row lengths are all half-integers,
and correspond to a ``Young tableau'' containing a column of ``half-boxes.''

\vs{.1in}
\noindent{\bf Integrable representations of $\soevennK$.} 
\vs{.1in}

\option
The Dynkin diagram for $\soevennK$ is

\begin{picture}(500,80)(10,10)
\put(100,40){\circle{5}}
\put(98,25){1}
\put(102,40){\line(1,0){26}}

\put(130,70){\circle{5}}
\put(120,66){0}
\put(130,68){\line(0,-1){26}}

\put(130,40){\circle{5}}
\put(128,25){2}
\put(132,40){\line(1,0){26}}

\put(160,40){\circle{5}}
\put(158,25){3}
\put(162,40){\line(1,0){26}}

\put(190,40){\circle{5}}
\put(188,25){4}
\put(192,40){\line(1,0){26}}

\put(220,40){\circle{5}}
\put(218,25){5}
\put(222,40){\line(1,0){26}}

\put(250,40){\circle{5}}
\put(248,25){6}
\put(252,40){\line(1,0){4}}
\put(258,40){\line(1,0){3}}
\put(263,40){\line(1,0){3}}
\put(268,40){\line(1,0){3}}
\put(273,40){\line(1,0){5}}

\put(280,40){\circle{5}}
\put(270,25){$n\!-\!3$}
\put(282,40){\line(1,0){26}}

\put(310,40){\circle{5}}
\put(301,25){$n\!-\!2$}
\put(312,40){\line(1,0){26}}

\put(310,70){\circle{5}}
\put(314,66){$n\!-\!1$}
\put(310,68){\line(0,-1){26}}

\put(340,40){\circle{5}}
\put(338,25){$n$}

\end{picture}
\option
and the dual Coxeter labels are 
($m_0$, $m_1$, $m_2, \cdots, m_{n-2}, m_{n-1}, m_{n}$)
= (1, 1, 2, $\cdots$, 2, 1, 1),
where the labelling of nodes is indicated on the diagram.
Integrable representations of $\soevennK$ thus have Dynkin indices 
that satisfy\footnote{
\noindent
For $\sofourK$,
we take the integrability condition to be
$ a_0 + \max( a_1, a_2) = K$,
which is equivalent to 
$\ell_1 + | \ell_2 | \le K$.
}
\be
\label{eq:soevenintegrable}
a_0 + a_1 + 2(a_2 + \cdots + a_{n-2}) + a_{n-1} + a_n = K\,.
\ee
An even or odd value of $a_n -a_{n-1}$ corresponds, 
respectively,
to a tensor or spinor representation of $\soevenn$.

The Dynkin diagram of $\soevenn$ (and also $\soevennK$)
is invariant under the exchange of the 
$(n-1)^{\rm th}$ and $n^{\rm th}$ nodes. 
This gives rise to a $\Z_2$-automorphism $\varepsilon$
of the $\soevenn$ Lie algebra, 
which exchanges representations with Dynkin indices
$(\cdots, a_{n-1}, a_{n})$
and 
$(\cdots, a_{n}, a_{n-1})$.
This automorphism may be dubbed \cite{Gaberdiel:2002qa}
``chirality flip'' 
as it exchanges the two fundamental 
spinor representations of opposite chirality.

For each representation of $\soevenn$ we may define
\be
\ell_i = \left\{  
\begin{array}{ll}
\smhalf (a_n + a_{n-1})  + \sum_{j=i}^{n-2} a_j 
&
{\rm for~} 1 \le i \le n-2,  
\\ [.05in]
\smhalf (a_n + a_{n-1})  
&
{\rm for~} i=n-1, 
\\[.05in]
\smhalf 
(a_n - a_{n-1})
&
{\rm for~} i=n .
\end{array}
\right.
\ee
in terms of which the integrability condition 
(\ref{eq:soevenintegrable}) 
becomes $\ell_1 + \ell_2 \le K$.
The {\it absolute values} of $\ell_i$ represent 
the row lengths of a Young tableau $A$ with up to $n$ rows.
(For spinor representations, these row lengths are all half-integer,
and correspond to a Young tableau containing a column of half-boxes.)
When $a_n = a_{n-1}$, the Young tableau has $n-1$ or fewer rows,
and corresponds to a unique irreducible $\soevenn$ representation $a$,
one which is invariant under $\varepsilon$.
When $a_n \neq a_{n-1}$, the Young tableau has precisely $n$ rows
and corresponds to two distinct representations, 
$a$ and $\varepsilon(a)$.
Hence we may consider the Young tableau $A$ as labelling
either an irreducible ($a$) 
or a reducible ($a\oplus \varepsilon(a)$)
representation of $\soevenn$,
depending respectively on whether 
the representation $a$ is or is not invariant under $\varepsilon$.
We thus write
\be
\label{eq:defA}
A =  2^{s(a)-1} \left[ a \oplus \varepsilon(a) \right]
\ee
where 
\be
\label{eq:defs}
s(a) =  \left\{   
\begin{array}{ll}
1 &
{\rm if~} \ell_n \neq 0; {\rm~~ that~is,~} a \neq \varepsilon(a),  \\ [.05in]
0 &
{\rm if~} \ell_n =  0; {\rm~~ that~is,~} a =  \varepsilon(a).
\end{array}
\right.
\ee
(For {\it all} representations of $\sooddn$,
we define $\varepsilon(a)=a$ and $s(a)= 0$.)

Let $S_{ab}$ denote the (symmetric) modular transformation matrix of $\soNK$,
an explicit formula for which may be found, for example,
in ref.~\cite{Mlawer:1990uv}.
We define
\bea
\label{eq:defSAb}
S_{Ab} &=& 2^{s(a)-1} \left[ S_{ab} + S_{\varepsilon(a)b}\right], \nonumber\\
S_{AB} &=& 2^{s(b)-1} \left[ S_{Ab} + S_{A\varepsilon(b)}\right].
\eea
Since the modular transformation matrix obeys 
\be
\label{eq:epseps}
S_{\varepsilon(a)b}  = S_{a\varepsilon(b)} 
\ee
it follows that 
\bea
\label{eq:SAbeps}
S_{Ab} &=& S_{A\varepsilon(b)},  \nonumber\\
S_{AB} &=& 2^{s(b)} S_{Ab}. 
\eea

\vs{.1in}
\noindent{\bf Simple current orbits of $\soNK$.} 
\vs{.1in}

\option
Both $\sooddnK$ and $\soevennK$ Dynkin diagrams have a $\Z_2$-symmetry that 
exchanges the $0^{\rm th}$ and $1^{\rm st}$ nodes.
This symmetry induces a simple-current symmetry 
(denoted by $\sig$)
of the $\soNK$ WZW model
that pairs integrable representations
related by $a_0 \leftrightarrow a_1$,
with the other Dynkin indices unchanged.\footnote{\noindent
Except for $\sofourK$, in which case $\sig$ acts by 
$a_0 \leftrightarrow \min(a_1,a_2)$.
Thus, if $a$ has Dynkin indices $(a_0,\, a_1, \,a_2)$
then $\sig(a)$ has Dynkin indices $(\min(a_1,a_2), \,K- a_2, \,K- a_1)$.
}
Their respective Young tableaux are related by 
$\ell_1 \to K - \ell_1$.
Under $\sig$, tensor representations are mapped to tensors, 
and spinor representations to spinors.

We will refer to representations of $\soNK$ with 
$\ell_1 < \smhalf K$,
$\ell_1 = \smhalf K$, and 
$\ell_1 > \smhalf K$ 
as being of types I, II, and III respectively.
Type II representations are invariant under $\sig$,
and are tensors (resp. spinors) when $K$ is even (resp. odd).
Each simple-current orbit 
of $\soNK$ contains either 
a type I and type III representation, 
or a single type II representation.
We define
\be
\label{eq:deft}
t(a) =  \left\{   
\begin{array}{ll}
1 &
{\rm if~} \ell_1  = \smhalf K ~({\rm type~II}); 
{\rm~~ that~is,~} a = \sig(a),  \\ [.05in]
0 &
{\rm if~} \ell_1 \neq \smhalf K ~({\rm type~I~or~III}); 
{\rm~~ that~is,~} a \neq \sig(a). 
\end{array}
\right.
\ee
Finally,
the modular transformation matrix of $\soNK$ obeys
\be
\label{eq:sigeps}
S_{\sig(a)b}  = \pm S_{\varepsilon(a)b}  \quad 
{\rm for~} b {\rm~~a~}
\left\{ {\rm tensor} \atop {\rm spinor} \right\}  {\rm ~representation.} 
\ee

\section{Level-rank duality of untwisted D-branes of $\soNK$}
\setcounter{equation}{0}
\label{secuntwisted}

Having reviewed the characterization of integrable 
representations of $\soNK$ in the previous section,  
we now turn to the untwisted D-branes of the $\soNK$ WZW model,
which are labelled by those representations.
In this section, we will demonstrate a level-rank duality
between the untwisted D-branes of $\soNK$ and those of $\soKN$.

Since the multiplicities of the representations carried by 
an open string stretched 
between two untwisted D-branes are given by the fusion coefficients 
of the WZW model (\ref{eq:verlinde}),
level-rank duality of the untwisted D-branes of the $\soNK$ model
is closely related to 
level-rank duality of the fusion coefficients of this model, 
which was described
in ref.~\cite{Mlawer:1990uv}.   
We recall two salient aspects of this duality:

\begin{itemize}
\item The level-rank map is partial:
it only relates the {\it tensor} representations\footnote{
For $\sooddnoddk$,
a level-rank map between the spinor representations 
also exists \cite{Mlawer:1990uv},
and thus a level-rank map can be defined for all the untwisted D-branes
of $\sooddnoddk$.
}
of $\soNK$ to those of $\soKN$.

\item The 
level-rank map is not one-to-one between integrable tensor representations $a$, 
but rather between {\it equivalence classes} of representations,\footnote{
\noindent
This is also the case for level-rank duality of $\suNK$.}
denoted by $[a]$.
These equivalence classes are characterized by 
tensor Young tableaux with 
$ \le\!\! N/2$ rows and $ \le\!\! K/2$ columns
(termed ``reduced and cominimally-reduced'' in ref.~\cite{Mlawer:1990uv}).
Level-rank duality acts by transposing these tableaux,
and maps the set of tensor Young tableaux 
with $ \le \!\!N/2$ rows and $ \le \!\!K/2$ columns
one-to-one onto the set of tensor Young tableaux 
with $ \le \!\!K/2$ rows and $ \le \!\!N/2$ columns.

\end{itemize}

\noindent
The equivalence classes 
of integrable tensor representations
fall into several categories, which we now describe,
using the notation of the previous section.

\vs{.1in}
\noindent (1)  $s(a) = 0$ and $t(a)= 1$:
the equivalence class labelled by 
a tensor Young tableau with
$\ell_1 = \half K$ columns 
(only possible when $K$ is even)
and with fewer than $\half N$ rows corresponds to 
a {\it single} (type II) irreducible representation $a$,
whose Dynkin indices satisfy
$a_0 = a_1$ 
and (for $N=2n$) $a_n = a_{n-1}$.
This representation is invariant under both $\sig$  and $\varepsilon$.

\vs{.1in}
\noindent (2)  $s(a) = 0$ and  $t(a)= 0$:
the equivalence class labelled by 
a tensor Young tableau with
$\ell_1 < \half K$ columns 
and with fewer than $\half N$ rows corresponds to 
a {\it pair} of irreducible representations 
$a$ and $\sig(a)$
(of type I and type III)
whose Dynkin indices 
are related by $a_0 \leftrightarrow a_1$.
When $N=2n$, 
the Dynkin indices of these representations satisfy $a_n = a_{n-1}$,
i.e.,  these representations are invariant under $\varepsilon$.

\vs{.1in}
\noindent (3)  $s(a) = 1$ and  $t(a)= 1$:
the equivalence class labelled by 
a tensor Young tableau with
$\ell_1 = \half K$ columns 
(only possible when $K$ is even)
and with exactly $\half N$ rows 
(only possible when $N$ is even)
corresponds to 
a {\it pair} of (type II) irreducible representations 
$a$ and $\varepsilon(a)$,
whose Dynkin indices 
are related by $a_n \leftrightarrow a_{n-1}$ where $N =2n$,
and obey\footnote{For $\sofourK$, they obey $a_0 = \min(a_1,a_2)$.}
$a_0 = a_1$,
i.e.,  these representations are invariant under $\sig$.

\vs{.1in}
\noindent (4)  $s(a) = 1$ and  $t(a)= 0$:
the equivalence class labelled by 
a tensor Young tableau with
$\ell_1 < \half K$ columns 
and with exactly $\half N$ rows 
(only possible when $N$ is even)
corresponds to 
{\it four} irreducible representations:
$a$, 
$\sig(a)$, 
$\varepsilon(a)$, 
and 
$\varepsilon(\sig(a))$
(two of type I and two of type III).

\vs{.1in}

\noindent
Let $[\tilde{a}]$ denote the transpose of the Young tableau 
characterizing the equivalence class $[a]$.
Then 
\be
\label{eq:defts}
t(a) = \tilde{s}(\tilde{a})  
\qquad {\rm and} \qquad
s(a) = \tilde{t}(\tilde{a})
\ee
where $\tilde{s}$ and $\tilde{t}$ are
the quantities (\ref{eq:defs}) and (\ref{eq:deft}) defined for $\soKN$.
Under level-rank duality, 
equivalence classes $[a]$ in categories (1), (2), (3), and (4) map into 
equivalence classes $[\tilde{a}]$ in categories (4), (2), (3), and (1) 
respectively. 

We now elucidate the implications of level-rank duality 
for the untwisted tensor D-branes of the $\soNK$ WZW model.
Consider the linear combination of untwisted Cardy states 
\be
\equiva
= 
{1 \over {\sqrt2}^{~t(\a) - s(\a) + 3 }}
\biggl[
  | \a \rangle\!\rangle_C
+ | \sig(\a) \rangle\!\rangle_C
+ | \varepsilon(\a)  \rangle\!\rangle_C
+ | \varepsilon(\sig(\a))  \rangle\!\rangle_C
\biggr]
\ee
which corresponds to an equivalence class $[\a]$ 
of integrable tensor representations.
Using eqs.~(\ref{eq:verlinde}) and (\ref{eq:sigeps}),
we find that the multiplicity $\sblama$ 
of the equivalence class of representations $[\lam]$ 
carried by an open string stretched between 
untwisted D-branes corresponding to the states 
$\equiva$
and 
$\equivb$
is given by 
\bea
\label{eq:defsblama}
\sblama 
&=&
{1 \over {\sqrt2}^{~t(\a)+t(\b)+t(\lam) - s(\a)-s(\b)-s(\lam)+3}  }
\times \nonumber\\
&&\qquad
\sum_{   \mu = {\rm tensor} \atop {\rm representations}  }
{ 
\left( S_{\b \mu} + S_{\varepsilon(\b)\mu} \right)
\left( S_{\lam \mu} + S_{\varepsilon(\lam)\mu} \right)
\left( S^*_{\a \mu} + S^*_{\varepsilon(\a)\mu} \right)
 \over S_{\id\mu} 
} 
\eea
where only integrable {\it tensor} representations $\mu$ remain in the sum
as a consequence of eq.~(\ref{eq:sigeps}).
Using eqs.~(\ref{eq:defSAb}) and (\ref{eq:SAbeps}),
we express $\sblama$  in terms of Young tableaux
$A$, $B$, $\Lam$, $M$, related to $\a$, $\b$, $\lam$, and $\mu$ by
eq.~(\ref{eq:defA}),
\be
\sblama 
=
{1 \over {\sqrt2}^{~t(\a)+t(\b)+t(\lam) + s(\a)+s(\b)+s(\lam)-3 }}
\sum_{   M = {\rm tensor} \atop {\rm tableaux~I,II,III}  }
{1 \over 2^{s(\mu)} }
{ 
S_{B M} 
S_{\Lam M} 
S^*_{A M} 
\over S_{\id M} 
}  \,.
\ee
Finally, since 
$S_{A \sig(M)} = S_{A M}$,
the sum may be restricted to tableaux of types I and II
(``cominimally-reduced'' tableaux)
\be
\label{eq:sblamafinal}
\sblama 
=
{1 \over {\sqrt2}^{~t(\a)+t(\b)+t(\lam) + s(\a)+s(\b)+s(\lam)-3 }}
\sum_{ M  = {\rm tensor}\atop {\rm tableaux~I,II} }
{1 \over 2^{s(\mu)+t(\mu)-1}  }
{ 
S_{B M} 
S_{\Lam M} 
S^*_{A M} 
 \over S_{\id M} 
} \,.
\ee
The multiplicities $\sblama$ are closely 
related to ${\Sigma_{B\Lam}}^A$ 
defined in eq.~(3.16) of ref.~\cite{Mlawer:1990uv}.

Level-rank duality maps the 
state $\equiva$ of $\soNK$ to the 
state $\equivta$ of $\soKN$.
Let  $\tilde{n}_{[\tb] [\tlam] }^{~~~~~[\ta]}  $
denote the quantity (\ref{eq:defsblama}) defined for $\soKN$.
The form of eq.~(\ref{eq:sblamafinal}) makes manifest the equality
of the multiplicities
\be
\sblama = \tilde{n}_{[\tb] [\tlam] }^{~~~~~[\ta]}  
\ee
as a consequence of three facts:
(1) the set of cominimally-reduced tableaux $M$ of $\soNK$ 
are in one-to-one correspondence with those of $\soKN$,
(2) eq.~(\ref{eq:defts}) holds for all tensor representations,
and
(3) the quantities $S_{AB}$, defined by eq.~(\ref{eq:defSAb}),
 are level-rank dual
($S_{AB} = \tS_{\tA \tB}$)
as was proved in the appendix of ref.~\cite{Mlawer:1990uv}.
Hence, the spectrum of representations carried by open strings
stretched between untwisted tensor D-branes of $\soNK$ is level-rank dual.

We end this section by describing the level-rank duality of 
untwisted spinor D-branes of $\sooddnoddk$.
The equivalence classes $[\a]$ of spinor representations of $\sooddnoddk$
are characterized by type I and type II spinor tableaux,
where a type I tableau represents a pair of spinor representations
$\a$ and $\sig(\a)$,
and a type II tableau represents a single irreducible spinor representation 
$\a$ that obeys $\sig(\a)= \a$.  
The level-rank map $[\a] \to [\ha]$
between equivalence classes of spinor representations
of $\sooddnoddk$ and $\sooddkoddn$ was presented in 
ref.~\cite{Mlawer:1990uv}:
\begin{itemize}
\item
reduce each of the row lengths by $\smhalf$,
so that they all become integers
\item
transpose the resulting tableau
\item
take the complement with respect to a $k \times n$ rectangle
and rotate 180 degrees
\item
add $\smhalf$ to each of the row lengths.
\end{itemize}
This takes type I spinor tableaux of $\sooddnoddk$ to type II spinor tableaux
of $\sooddkoddn$ and vice versa:
$t(\a) = 1 - \tilde{t}(\ha)$.
This procedure thus defines a map between an untwisted 
spinor D-brane of $\sooddnoddk$ corresponding to the boundary state
\be
\equiva
= 
{1 \over {\sqrt2}^{~t(\a) + 1 }}
\biggl[
  | \a \rangle\!\rangle_C
+ | \sig(\a) \rangle\!\rangle_C
\biggr]
\ee
and an untwisted spinor D-brane $\equivha$ of $\sooddkoddn$.
The multiplicity
of the (equivalence class of) representations $[\lam]$ 
carried by an open string stretched
between untwisted spinor D-branes $[\a]$ and $[\b]$ of $\sooddnoddk$
obeys
\be
\sblama 
= {1 \over {\sqrt2}^{~t(\a)+t(\b)-5 }}
\sum_{ \mu  = {\rm tensor}\atop {\rm tableaux~I} }
{ 
S_{\b \mu} 
S_{\lam \mu} 
S^*_{\a \mu} 
 \over S_{\id \mu} 
} 
= \tilde{n}_{[\hb] [\tlam] }^{~~~~~[\ha]}  
\ee
using eq.~(3.25) of ref.~\cite{Mlawer:1990uv}.
Hence, the spectrum of representations carried by open strings
stretched between untwisted spinor D-branes of $\sooddnoddk$ is 
also level-rank dual.

\section{The $\varepsilon$-twisted D-branes of the $\soevennK$ model} 
\setcounter{equation}{0}
\label{sectwisted}

In the previous section, we proved the level-rank duality of
untwisted D-branes of $\soNK$.
In this section, we will describe a class of 
twisted D-branes of $\soevennK$, and in the next section, 
we will prove the level-rank duality of 
a subset of these twisted D-branes.

Recall from in sec.~\ref{secsoNK}
that the finite Lie algebra $\soevenn$ 
possesses (when $n \ge 2$) 
a $\Z_2$-automorphism $\varepsilon$ (chirality flip),
under which the Dynkin indices $a_{n-1}$ and $a_n$ 
of an irreducible representation are exchanged.
This automorphism lifts to an automorphism of 
the affine Lie algebra $\soevennK$,
and 
gives rise to a set of $\varepsilon$-twisted Ishibashi states
and $\varepsilon$-twisted Cardy states of the bulk $\soevennK$ 
WZW model,
and a corresponding class of $\varepsilon$-twisted 
D-branes of the boundary model.
In this section we characterize these twisted states,
relying heavily on ref.~\cite{Gaberdiel:2002qa}.

\vs{.1in}
\noindent{\bf $\varepsilon$-twisted Ishibashi states of $\soevennK$.}
\vs{.1in}

\option 
The $\varepsilon$-twisted Ishibashi states $\epsishimu$ of 
the $\soevennK$ WZW model
are labelled by integrable representations 
$\mu \in {\cal E}^{\varepsilon} $
of $\soevennK = \DnoneK$
that obey $\varepsilon(\mu) = \mu$
(i.e., integrable representations
characterized by $\soevenn$ tensor Young tableaux 
with no more than $n-1$ rows).
These representations have Dynkin indices 
\be
\label{eq:reducedtensor}
(\mu_0, \mu_1,  \cdots,  \mu_{n-2}, \mu_{n-1}, \mu_{n-1})
\ee
that satisfy\footnote{For $\sofourK$, 
the Dynkin indices $(\mu_0, \mu_1,  \mu_1)$ satisfy
$\mu_0 + \mu_1= K$.}
\be
\label{eq:integrableIshieven}
\mu_0 + \mu_1 + 2(\mu_2 + \cdots + \mu_{n-1})  = K \,.
\ee
Equivalently, the $\varepsilon$-twisted Ishibashi states of $\soevennK$ may be
characterized by the  
integrable representations of the associated orbit Lie algebra 
$\gcup^\varepsilon = (A_{2n-3}^{(2)})_{K}$ \cite{Gaberdiel:2002qa}
whose Dynkin diagram is

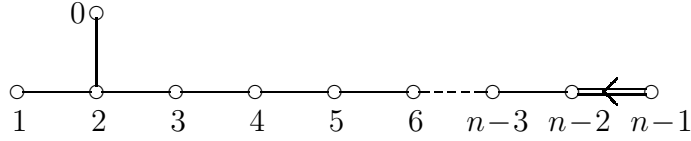
\begin{picture}(500,80)(10,10)
\put(100,40){\circle{5}}
\put(98,25){1}
\put(102,40){\line(1,0){26}}

\put(130,70){\circle{5}}
\put(120,66){0}
\put(130,68){\line(0,-1){26}}

\put(130,40){\circle{5}}
\put(128,25){2}
\put(132,40){\line(1,0){26}}

\put(160,40){\circle{5}}
\put(158,25){3}
\put(162,40){\line(1,0){26}}

\put(190,40){\circle{5}}
\put(188,25){4}
\put(192,40){\line(1,0){26}}

\put(220,40){\circle{5}}
\put(218,25){5}
\put(222,40){\line(1,0){26}}

\put(250,40){\circle{5}}
\put(248,25){6}
\put(252,40){\line(1,0){4}}
\put(258,40){\line(1,0){3}}
\put(263,40){\line(1,0){3}}
\put(268,40){\line(1,0){3}}
\put(273,40){\line(1,0){5}}

\put(280,40){\circle{5}}
\put(270,25){$n\!-\!3$}
\put(282,40){\line(1,0){26}}

\put(310,40){\circle{5}}
\put(301,25){$n\!-\!2$}
\put(312,41){\line(1,0){26}}
\drawline(322,40)(327,45)
\drawline(322,40)(327,35)
\put(312,39){\line(1,0){26}}

\put(340,40){\circle{5}}
\put(332,25){$n\!-\!1$}

\end{picture}

\option
and whose dual Coxeter numbers are
($m_0$, $m_1$, $m_2, \cdots, m_{n-1}$)
= (1, 1, 2, $\cdots$, 2),
where the labelling of nodes is indicated on the diagram.
The $\DnoneK$ representation 
with Dynkin indices
(\ref{eq:reducedtensor})
corresponds to the 
$ (A_{2n-3}^{(2)})_{K}$
representation with Dynkin indices
($\mu_0$, $\mu_1, \cdots,  \mu_{n-2}$, $\mu_{n-1}$),
whose integrability condition is precisely (\ref{eq:integrableIshieven}).

It was shown in ref.~\cite{Gaberdiel:2002qa}
that each $\varepsilon$-twisted Ishibashi state $\mu$ of $\soevennK$ 
may be mapped 
to a {\it spinor} representation $\mu'$ 
of the untwisted affine Lie algebra $\somp$ 
with Dynkin indices\footnote{Except for 
$\sofourK$, in which case 
$\mu'_0= 2\mu_0 +1$ and 
$\mu'_1 =2\mu_1 + 1$.
Also, recall footnote \ref{fn}.}
\be
\label{eq:defmuprime}
\mu'_i = \mu_i\quad (0 \le i \le n-2)
\qquad{\rm and}\qquad
\mu'_{n-1} =  2\mu_{n-1} + 1 .
\ee
The constraint (\ref{eq:integrableIshieven}) on $\mu$ 
is precisely equivalent to 
the integrability constraint (\ref{eq:sooddintegrable}) on 
the representation $\mu'$ of $\somp$.
Hence, 
{\it $\varepsilon$-twisted Ishibashi states of $\soevennK$
are in one-to-one correspondence with the
set of integrable spinor representations of $\somp$
of type I, type II (when $K$ is even), and type III.}

\vs{.2in}
\noindent{\bf $\varepsilon$-twisted Cardy states of $\soevennK$.}
\vs{.1in}

\option 
The $\varepsilon$-twisted Cardy states $\epscardya$
(and therefore the $\varepsilon$-twisted D-branes) 
of the $\soevennK$ WZW model\footnote{
$n \ge 2$ is understood.}
are labelled 
by the integrable representations 
$\a \in \Boundeps$ 
of the $\varepsilon$-twisted affine Lie algebra 
${\hat{g}_K^\varepsilon} = \DntwoK$ \cite{Gaberdiel:2002qa},
whose Dynkin diagram is 

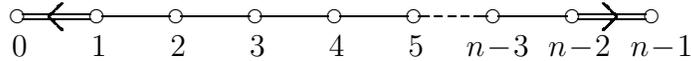
\begin{picture}(500,50)(10,10)
\put(100,40){\circle{5}}
\put(98,25){0}
\put(102,39){\line(1,0){26}}
\drawline(112,40)(117,45)
\drawline(112,40)(117,35)
\put(102,41){\line(1,0){26}}

\put(130,40){\circle{5}}
\put(128,25){1}
\put(132,40){\line(1,0){26}}

\put(160,40){\circle{5}}
\put(158,25){2}
\put(162,40){\line(1,0){26}}

\put(190,40){\circle{5}}
\put(188,25){3}
\put(192,40){\line(1,0){26}}

\put(220,40){\circle{5}}
\put(218,25){4}
\put(222,40){\line(1,0){26}}

\put(250,40){\circle{5}}
\put(248,25){5}
\put(252,40){\line(1,0){4}}
\put(258,40){\line(1,0){3}}
\put(263,40){\line(1,0){3}}
\put(268,40){\line(1,0){3}}
\put(273,40){\line(1,0){5}}

\put(280,40){\circle{5}}
\put(270,25){$n\!-\!3$}
\put(282,40){\line(1,0){26}}

\put(310,40){\circle{5}}
\put(301,25){$n\!-\!2$}
\put(312,41){\line(1,0){26}}
\drawline(327,40)(322,45)
\drawline(327,40)(322,35)
\put(312,39){\line(1,0){26}}

\put(340,40){\circle{5}}
\put(332,25){$n\!-\!1$}

\end{picture}

\option
and whose dual Coxeter numbers are
($m_0$, $m_1, \cdots,m_{n-2}$, $m_{n-1}$)
= (1, 2, $\cdots$, 2, 1),
where the labelling of nodes is indicated on the diagram.
The Dynkin indices 
$(\a_0, \a_1, \cdots, \a_{n-2}, \a_{n-1})$ of $\a$  
thus satisfy\footnote{For $\sofourK$, the condition 
is $\a_0 + \a_1 = K$.}
\be
\label{eq:integrableCardyeven}
\a_0 + 2( \a_1 + \cdots + \a_{n-2}) + \a_{n-1} = K\,.
\ee

It was shown in ref.~\cite{Gaberdiel:2002qa}
that each $\varepsilon$-twisted Cardy state $\a$ of $\soevennK$
may be mapped to a representation $\a'$ 
of the untwisted affine Lie algebra $\somp$ 
with Dynkin indices\footnote{Except for 
$\sofourK$, in which case 
$\a'_0= 2\a_0 + \a_1 + 2$ and $\a'_1 = \a_1$.
Also, recall footnote \ref{fn}.}
\be
\label{eq:defaprime}
\a'_0 = \a_0 + \a_1 + 1 \qquad{\rm and}\qquad
\a'_i = \a_i \quad  (1 \le i \le n-1) .
\ee
The constraint (\ref{eq:integrableCardyeven})  on $\a$ 
implies that $\a'$ is an integrable representation of $\somp$,
and the constraint $\a_0 \ge 0$ further implies that
$\a'$ is a type I representation of $\somp$ 
(i.e., corresponds to a Young tableau
whose first row length obeys $\ell_1'  \le \half K$).
Therefore, 
{\it $\varepsilon$-twisted D-branes of $\soevennK$ 
are in one-to-one correspondence with the
set of integrable type I representations of $\somp$,
of both tensor and spinor types.}

Although the $\varepsilon$-twisted Ishibashi states 
and the $\varepsilon$-twisted Cardy  states of $\soevennK$
are characterized differently 
in terms of integrable representations of $\somp$,
they are equal in number.
(For $K$ even, the number of each is 
$\left( n - 1 + K/2  \atop n-1 \right)
+\left( n - 2 + K/2  \atop n-1 \right)$,
while for $K$ odd, the number of each is
$ 2 \left( n - 1 + (K-1)/2  \atop n-1 \right)$.)
Thus, the $\varepsilon$-twisted Cardy states $\a$  may be written 
as linear combinations (\ref{eq:defpsi})
of $\varepsilon$-twisted Ishibashi states $\mu$,
with the transformation coefficients $\psi_{\a\mu}$ 
given by the modular transformation matrix of $\DntwoK$.
In ref.~\cite{Gaberdiel:2002qa}, it was shown that
these coefficients are proportional to the
(real) matrix elements $ S'_{\a'\mu'}$
of the modular transformation matrix  
of the untwisted affine Lie algebra $\somp$:
\be
\label{eq:psi}
\psi_{\a\mu} =   \sqrt{2}  S'_{\a'\mu'}
             =   \sqrt{2}  S^{\prime*}_{\a'\mu'}
\ee
where $\a'$ and $\mu'$ are the $\somp$ representations
related to $\a$ and $\mu$ by 
eqs.~(\ref{eq:defaprime}) and (\ref{eq:defmuprime}) respectively. 

\vs{.1in}
\noindent{\bf Twisted open string partition function of $\soevennK$.}
\vs{.1in}

\option
Combining eqs.~(\ref{eq:defn}) and (\ref{eq:psi}),
we may write the multiplicities of the representations carried by 
an open string
stretched between $\varepsilon$-twisted D-branes  $\a$ and $\b$
of $\soevennK$  as
\be
\label{eq:soevencoeff}
\nblama 
=  \sum_{ \mu' = {\rm spinors~I,II,III} }
{2\; \Spr_{\a'\mu'} S_{\lam\mu} \Spr_{\b'\mu'} 
\over S_{\id\mu} }  
\ee  
where
$S_{\lam\mu} $ and $S'_{\a'\mu'}$ 
are modular transformation matrix elements 
of $\soevennK$ and $\somp$ respectively,
and the sum is over all $\varepsilon$-twisted Ishibashi states $\mu$
of $\soevennK$, or equivalently, 
over all integrable spinor representations $\mu'$
of $\somp$.
(Type II spinors of $\somp$ are present only when $K$ is even.)

Although the $\varepsilon$-twisted D-branes 
correspond to both tensor and spinor representations of $\somp$,
for the remainder of this section 
we will restrict $\a$ and $\b$ to correspond to 
spinor representations $\a'$ and $\b'$ of $\somp$,
which allows us to simplify eq.~(\ref{eq:soevencoeff}) considerably.
Recall from eq.~(\ref{eq:sigeps})
that the modular transformation matrix elements of $\somp$ obey
\be
S'_{\a'\sig(\mu')}  = - S'_{\a'\mu'}  \qquad{\rm for~}\a' = {\rm~spinor}.
\ee
As a consequence, 
type I and III representations $\mu'$,
which are related by $\sig$,
may be combined in eq.~(\ref{eq:soevencoeff})
\be
\nblama 
=  \sum_{ \mu' = {\rm spinors~I} }
\left[ {2\; \Spr_{\a'\mu'} S_{\lam\mu} \Spr_{\b'\mu'} 
\over S_{\id\mu} }  
~  +  ~
     {2\; \Spr_{\a'\mu'} S_{\lam\sig(\mu)} \Spr_{\b'\mu'} 
\over S_{\id\sig(\mu)} }  
\right] 
\qquad {\rm for~}\a', \b' {\rm~both~spinors}
\ee
and type II representations, 
which obey $\sig(\mu') = \mu'$,
drop out of the sum since $S'_{\a'\mu'}  = 0$. 
(We have also used the fact that the $\somp$ representation $\sig(\mu')$,
related to $\mu'$ by $\ell'_1 \to K + 1 - \ell'_1$,
corresponds via the map (\ref{eq:defmuprime})
to the $\varepsilon$-twisted Ishibashi state $\sig(\mu)$ of $\soevennK$,
related to $\mu$ by $\ell_1 \to K - \ell_1$.)
Finally, recalling that the modular transformation matrix elements
of $\soevennK$ obey
\be
S_{\lam \sig(\mu)}  = \pm S_{\lam \varepsilon(\mu)}  \quad 
{\rm for~} \lam {\rm~a~}
\left\{ {\rm tensor} \atop {\rm spinor} \right\}  {\rm ~representation} 
\ee
and that $\varepsilon$-twisted Ishibashi states obey $\varepsilon(\mu) = \mu$,
we finally obtain
\be
\nblama =  
\sum_{ \mu' = {\rm spinors~I} }
{4\; \Spr_{\a'\mu'} S_{\lam\mu} \Spr_{\b'\mu'} 
\over S_{\id\mu} }   
 \qquad { {\rm for~} \a', \b' {\rm~both~spinors} \atop
           {\rm and~} \lam  = {\rm~tensor}  }
\ee
and 
\be
\label{eq:null}
\nblama =  
0  
\qquad \qquad\qquad  \qquad\qquad  \qquad 
{ {\rm for~} \a', \b' {\rm~both~spinors} \atop
                        {\rm and~} \lam  = {\rm~spinor}.  }
\ee
This result will allow us to demonstrate in the next section 
the level-rank duality of the spectrum of an open string 
stretched between spinor $\varepsilon$-twisted D-branes of $\soevennevenk$.

\section{Level-rank duality of 
$\varepsilon$-twisted D-branes of $\soevennevenk$}
\setcounter{equation}{0}
\label{sectwistedduality}

As we saw in the previous section,
the $\soevennK$ WZW model possesses twisted D-branes
corresponding to the chirality-flip symmetry $\varepsilon$ of the 
$\soevenn$ Dynkin diagram,
and these  $\varepsilon$-twisted D-branes are characterized
by integrable type I tensor and spinor representations of $\somp$.
We will refer to these as tensor and spinor $\varepsilon$-twisted
D-branes respectively.

In this section, we will exhibit a level-rank duality\footnote{
\noindent
Clearly the  $\varepsilon$-twisted D-branes of $\soevennoddk$ 
have no level-rank duals, since $\sooddkevenn$ has no
$\varepsilon$-twisted D-branes.}
between the $\varepsilon$-twisted D-branes of $\soevennevenk$ 
and those of $\soevenkevenn$.
This duality is partial, 
and only holds between {\it spinor} $\varepsilon$-twisted D-branes
(just as the level-rank duality of untwisted D-branes 
only holds between tensor D-branes).
The restriction to spinor $\varepsilon$-twisted D-branes  
can be anticipated by observing that 
the number of tensor $\varepsilon$-twisted D-branes of $\soevennevenk$ is 
$ \left(  n+k-1 \atop n-1 \right) $
and the number of spinor $\varepsilon$-twisted D-branes is 
$ \left(  n+k-2 \atop n-1 \right) $,
and only the latter is invariant under $n \leftrightarrow  k$.

First we define an explicit one-to-one map $\a \to \ha$ 
between the spinor 
$\varepsilon$-twisted D-branes of $\soevennevenk$ and $\soevenkevenn$.
The map $\a \to \ha$ 
is defined by specifying its action\footnote{
Note that the ``hat'' map defined here differs from that
defined in sec.~\ref{secuntwisted} between spinor 
representations of $\sooddnoddk$ and $\sooddkoddn$.
The map defined here also characterizes the 
map between $\om_c$-twisted D-branes of
${\widehat{\rm su}(2n+1)_{2k+1}}$
and ${\widehat{\rm su}(2k+1)_{2n+1}}$ 
\cite{Naculich:2006bf}.}
on the corresponding $\sonmkp$ and $\sokmnp$ 
representations $\a'$ and $\ha'$:

\begin{itemize}

\item  reduce each of the row lengths of $\a'$ by $\smhalf$,
so that they all become integers
\item transpose the resulting tableau
\item add $\smhalf$ to each of the row lengths.

\end{itemize}
\noindent 

The same procedure defines a one-to-one map $\mu' \to \hmu'$ 
between type I spinor representations of $\sonmkp$ and $\sokmnp$
corresponding to $\varepsilon$-twisted Ishibashi states.  
By virtue of eq.~(\ref{eq:defmuprime}), 
this map lifts to a map 
$\mu \to \tmu$
between (a subset of) 
the $\varepsilon$-twisted Ishibashi states.
As suggested by the notation, this map 
is simply transposition of the type I $\soevennevenk$
Young tableau corresponding to $\mu$. 

Next,
we turn to the level-rank duality of the spectrum of 
an open string stretched between $\varepsilon$-twisted D-branes.
In the previous section, 
it was shown that
the multiplicity of the representation $\lambda$
carried by an open string stretched between spinor
$\varepsilon$-twisted D-branes $\a$ and $\b$ of $\soevennevenk$
is  given by 
\be
\label{eq:nblamatensor}
\nblama =  
\sum_{ \mu' = {\rm spinors~I} }
{4\; \Spr_{\a'\mu'} S_{\lam\mu} \Spr_{\b'\mu'} 
\over S_{\id\mu} }   \qquad {\rm for~}\lam = {\rm~tensor}
\ee
with $\nblama$ vanishing for $\lam=$ spinor.
As in sec.~\ref{secuntwisted},  however,
we consider the multiplicity corresponding to the 
equivalence class of tensor representations $[\lam]$:
\be
\label{eq:nbLamadef}
\nbLama =
{1 \over {\sqrt2}^{~t(\lam) - s(\lam) + 3 }}
\left[  {n_{\b\lam}}^{\a}  +
        {n_{\b\varepsilon(\lam)}}^{\a}  +
        {n_{\b\sig(\lam)}}^{\a}  +
        {n_{\b\varepsilon(\sig(\lam))}}^{\a}  \right] .
\ee
Using eqs.~(\ref{eq:nblamatensor}), (\ref{eq:sigeps}), and (\ref{eq:defSAb}),
we find
\bea
\label{eq:nbLamafinal}
\nbLama &=&
{1 \over {\sqrt2}^{~t(\lam) - s(\lam) + 1 }}
\sum_{ \mu' = {\rm spinors~I} }
{4\; \Spr_{\a'\mu'} \left( S_{\lam\mu} 
                          +S_{\varepsilon(\lam)\mu}  \right) \Spr_{\b'\mu'} 
\over S_{\id\mu} }   \nonumber\\
&=&
{1 \over {\sqrt2}^{~t(\lam) + s(\lam) - 1 }}
\sum_{ \mu' = {\rm spinors~I} }
{4\; \Spr_{\a'\mu'} S_{\Lam M } \Spr_{\b'\mu'} 
\over S_{\id M } }   
\eea
where 
$\Lam =  2^{s(\lam)-1} \left[ \lam \oplus \varepsilon(\lam) \right]$
and $ M  = \mu$ since $\varepsilon(\mu) = \mu$.
The form of eq.~(\ref{eq:nbLamafinal}) makes manifest the equality
of the multiplicities 
\bea
\nbLama 
&= &
{1 \over {\sqrt2}^{~t(\lam) + s(\lam) - 1 }}
\sum_{ \mu' = {\rm spinors~I} }
{4\; \Spr_{\a'\mu'} S_{\Lam M } \Spr_{\b'\mu'} 
\over S_{\id M } }    \\
& = & 
{1 \over {\sqrt2}^{~\tilde{s}(\tlam) + \tilde{t}(\tlam) - 1 }}
\sum_{ \hmu' = {\rm spinors~I} }
{4\; \tSpr_{\ha'\hmu'} \tS_{\tLam\tM} \tSpr_{\hb'\hmu'} 
\over \tS_{\id\tM} }   \\
&=&
\tn_{\hb[\tlam]}^{~~~~\ha} 
\eea
where we have used eq.~(\ref{eq:defts}) and the facts that:

\noindent  (1) type I spinors  $\mu'$ of $\sonmkp$ map one-to-one to 
type I spinors $\hmu'$ of $\sokmnp$,

\noindent  (2) 
$S_{\Lam M } = \tS_{\tLam\tM}$ ~\cite{Mlawer:1990uv},
where $S$ and $\tS$ are the modular transformation matrices of
$\soevennevenk$ and $\soevenkevenn$ respectively,
and 

\noindent  (3) 
$ S'_{\a'\mu'}  = \tS'_{\ha'\hmu'} $
for $\a'$ and $\mu'$ both type I spinor representations
(eq.~(6.10) of ref.~\cite{Naculich:2006bf}), 
where $S'$ and $\tS'$ are the modular transformation matrices of
$\sonmkp$ and $\sokmnp$ respectively.
Since by eq.~(\ref{eq:null}) only tensor representations 
$\lam$ appear in the 
$\varepsilon$-twisted open-string partition function (\ref{eq:openpartition}),
we have established that the
spectrum of representations carried by open strings stretched between
$\varepsilon$-twisted D-branes of $\soevennevenk$ is level-rank dual.

\section{Conclusions}
\setcounter{equation}{0}
\label{secconcl}

We have analyzed the level-rank duality 
of the untwisted D-branes of $\soNK$ and 
of the $\varepsilon$-twisted D-branes of $\soevennevenk$.
In each case, only a subset of the D-branes are mapped
onto those of the level-rank-dual theory.

Untwisted D-branes of $\soNK$ are characterized by 
integrable tensor and spinor representations of $\soNK$.
Only the untwisted {\it tensor} D-branes participate in level-rank
duality.\footnote{
Except for $\sooddnoddk$, 
where the untwisted spinor D-branes also respect level-rank duality.}
The tensor representations $\a$  of $\soNK$ fall into equivalence classes
$[\a]$ generated by the $\Z_2$-isomorphisms $\sig$ and $\varepsilon$
(the latter non-trivial only for $N$ even),
and characterized by Young tableaux with 
$ \le\!\! N/2$ rows and $ \le\!\! K/2$ columns.
Level-rank duality acts by transposing these tableaux,
and thus maps the equivalence classes $[\a]$ of untwisted 
tensor D-branes of $\soNK$ onto $[\ta]$ of $\soKN$.
We showed that the multiplicity  $\sblama$
of the (equivalence class of) representations $[\lam]$
carried by an open string stretched between untwisted
$\soNK$ D-branes corresponding to $[\a]$ and $[\b]$
is equal to $\tilde{n}_{[\tb] [\tlam] }^{~~~~~[\ta]}  $,
the multiplicity  
of the (equivalence class of) representations $[\tlam]$
carried by an open string stretched between untwisted
$\soKN$ D-branes corresponding to $[\ta]$ and $[\tb]$.
A similar result was shown for untwisted spinor D-branes
of $\sooddnoddk$.

The $\varepsilon$-twisted D-branes of $\soevennevenk$,
associated with the chirality-flip symmetry $\varepsilon$ of the
$\soevenn$ Dynkin diagram,
are characterized by type I integrable 
tensor and spinor representations of $\sonmkp$.
Only the {\it spinor} $\varepsilon$-twisted D-branes 
participate in level-rank duality.
We defined a one-to-one map $\a \to \ha$ between 
the spinor $\varepsilon$-twisted D-branes of $\soevennevenk$ and 
the spinor $\varepsilon$-twisted D-branes of $\soevenkevenn$.
We then showed that the multiplicity  $\nbLama$
of the (equivalence class of) representations $[\lam]$
carried by an open string stretched between 
$\varepsilon$-twisted $\soevennevenk$ D-branes 
corresponding to $\a$ and $\b$
is equal to 
$\tn_{\hb[\tlam]}^{~~~~\ha} $,
the multiplicity  
of the (equivalence class of) representations $[\tlam]$
carried by an open string stretched between 
$\varepsilon$-twisted $\soevenkevenn$ D-branes 
corresponding to $\ha$ and $\hb $.

Hence, for both untwisted and $\varepsilon$-twisted D-branes,
we have established an isomorphism between the spectrum of an open string 
ending on these D-branes and 
the spectrum of an open string ending on the level-rank-dual D-branes.

In both the $\suNK$ and $\spnk$ WZW theories,
the charges of level-rank-dual untwisted D-branes are equal (modulo sign)
\cite{Naculich:2005tn,Naculich:2006mt},
with a slightly more complicated relationship holding between
the charges of twisted D-branes \cite{Naculich:2006bf}.
In the case of $\soNK$, however,
the charges of the D-branes 
do not exhibit any simple relationship under level-rank duality.

\section*{Acknowledgments}

SN wishes to thank Howard Schnitzer for his collaboration 
on a long series of papers on which the results 
of this paper depend.

\providecommand{\href}[2]{#2}\begingroup\raggedright\endgroup
\end{document}